# Computational Approaches for Predicting Drug-Disease Associations: A Comprehensive Review

Chunyan Ao[1,2], Zhichao Xiao[1], Lixin Guan[1], Liang Yu[1*]

1. School of Computer Science and Technology, Xidian University, Xi'an, China
2. Institute of Fundamental and Frontier Sciences, University of Electronic Science and Technology of China, Chengdu, China

*Corresponding authors: lyu@xidian.edu.cn.

## Abstract

In recent decades, traditional drug research and development have been facing challenges such as high cost, long timelines, and high risks. To address these issues, many computational approaches have been suggested for predicting the relationship between drugs and diseases through drug repositioning, aiming to reduce the cost, development cycle, and risks associated with developing new drugs. Researchers have explored different computational methods to predict drug-disease associations, including drug side effects-disease associations, drug-target associations, and miRNA-disease associations. In this comprehensive review, we focus on recent advances in predicting drug-disease association methods for drug repositioning. We first categorize these methods into several groups, including neural network-based algorithms, matrix-based algorithms, recommendation algorithms, link-based reasoning algorithms, and text mining and semantic reasoning. Then, we compare the prediction performance of existing drug-disease association prediction algorithms. Lastly, we delve into the present challenges and future prospects concerning drug-disease associations.

# 1. Introduction

Although pharmaceutical companies have heavily invested in new drug R&D technologies over the past few decades, productivity in terms of the number of new drugs approved per dollar spent and the quantity of initial investigational new drugs (INDs) has actually declined since the mid-1990s [1]. The process of developing a new drug is an immensely costly undertaking, with expenses typically falling within the range of 2 to 3 billion US dollars. Moreover, this endeavor is known to be time-consuming, often taking a minimum of 13 to 15 years to reach completion [2]. Additionally, 90% of the drug candidates presented for evaluation to the US Food and Drug Administration (FDA) fail to gain approval, further hindering their application in actual treatment [3]. Despite significant progress in technology and substantial investments in research and development, the number of newly approved drugs has remained stagnant. Furthermore, the Contract Research Organization (CRO) penetration rate of drugs continues to increase each year. The CRO penetration rate of a drug refers to the ratio of the annual human demand for new drugs to the actual output of new drugs [4]. Therefore, drug research and development remain crucial global issues.

Given the time, money and clinical trials required in traditional drug discovery, researchers and the pharmaceutical industry urgently need to find a cost-effective drug discovery strategy that overcomes these challenges. As a result, drug repositioning has garnered considerable interest from researchers and the pharmaceutical industry alike. Drug repositioning, alternatively referred to as drug reassignment, drug repurposing, therapeutic switching, drug redirection, or drug reprofiling [3]. Drug repositioning, also known as drug repurposing, is a valuable approach to discover novel indications for already existing drugs [5], leveraging their established safety and pharmacokinetic profiles [3] and are characterized by efficiency, low cost, and no risk [5]. Therefore, the use of drug repositioning strategies can not only shorten development time but also reduce R&D costs and risks. Moreover, the use of drug repositioning methods has also broken through the cost bottleneck in many countries, providing opportunities for developing drugs at lower investments [5].

In recent years, the observation of therapeutic effects of certain drugs on multiple diseases and the identification of specific side effects that could be beneficial for other diseases have further fueled drug repositioning efforts.[6] Researchers have started exploring the potential of existing drugs in treating additional diseases based on their broad-spectrum efficacy and side effect profiles. To facilitate this process and narrow down the number of potential drug-disease interactions for further experimental verification, computational methods have emerged as valuable tools. These methods help improve experimental efficiency, reduce costs, and provide insights into potential drug-disease associations.[7] After more than a decade of advancing machine learning techniques, harnessing their super learning ability to discover potential drug-disease interactions [8]. Consequently, the utilization of computational methods to predict drug-disease associations has been on the rise as well. The field of drug-disease association prediction has experienced remarkable advancements, including the integration of heterogeneous data sources [9], network-based approaches [10], machine learning [11] and deep learning techniques [12], the integration of multi-omics data [13], knowledge graph-based approaches [14], and the application of natural language processing (NLP) [15]. These developments have significantly improved our ability to predict and discover unknown drug-disease associations, facilitating the identification of potential therapeutic effects for existing drugs and accelerating the drug repositioning process. By leveraging computational methods, researchers can enhance experimental efficiency, reduce costs, and gain valuable understanding of the intricate relationships between drugs and diseases, thereby driving improvements in drug development.

In this review, we will present a comprehensive understanding of drug-disease association prediction, including its development and advancements in the field. We will explore various algorithmic approaches employed in drug-disease associations prediction, such as neural network-based algorithms, matrix-based algorithms, recommendation algorithms, and link-based reasoning algorithms. Furthermore, we will discuss text mining and semantic reasoning methods and their application in drug repositioning. By summarizing the advantages and disadvantages of each method and comparing their performance, we aim to provide insights into the current landscape of drug repositioning research. Overall, this review will shed light on predicting drug-disease association, its challenges, and the computational methods employed to expedite the discovery of potential drug-disease associations.

## 2. Classification of Drug-Disease Association Prediction Methods

In this section, we introduce current popular algorithms for predicting drug-disease interactions and categorize them into four groups: neural network algorithms, matrix-based prediction algorithms, recommendation algorithms, and algorithms built on text analytics and language intelligence. Table 1 presents a summary of all the methods discussed in this paper.

Table 1 Summary of the drug repositioning algorithm

| Method | Strategy | Input | Prediction out | Advantage | Disadvantage |
|---|---|---|---|---|---|
| GIPAE | FC layer<br>Random forest | Fingerprint<br>Drug Gaussian similarity<br>Disease Gaussian similarity<br>Disease semantic similarity | D-D interactions | Low running times<br>does not need 3D structures | complex feature representation |
| SKCNN | CNN<br>Random forest | Drug sigmoid kernel similarity<br>Drug structure similarity<br>Disease semantic similarity<br>Disease sigmoid kernel similarity | D-D interactions | High accuracy<br>does not need 3D structures | complex feature representation |
| SAEROF | Sparse autoencoder<br>Random forest | Drug structure similarity<br>Drug Gaussian similarity<br>Disease Gaussian similarity<br>Disease semantic similarity | D-D interactions | does not need 3D structures | Low accuracy<br>complex feature representation |
| GFPred | graph convolutional autoencoder<br>FC autoencoder<br>attention mechanism | Drug attributes<br>Drug similarity<br>Disease similarity<br>D-D association | D-D interactions | does not need 3D structures | Low speed |
| DRRS | SVT algorithm | Drug similarity<br>Disease similarity<br>D-D association | D-D interactions | Low running times<br>does not need 3D structures | Input sparsity affects performance<br>classification only |
| $DNL_{2,1}$-CMF | dual-network $L_{2,1}$-collaborative matrix factorization | Drug similarity<br>Disease similarity | D-D interactions | Low running times<br>does not need 3D structures | classification only |
| CMFMTL | Multi-Task Learning Collective Matrix Factorization | D-D association | D-D interactions | does not need 3D structures | classification only |
| DRCFFS | collaborative filtering | drug chemical structures drug target proteins D-D association | D-D interactions | Low running times High accuracy | classification only |
| MeSHDD | bit-wise distance robust clusters | MEDLINE repository terms | D-D pairs | Text input only | No gold standard testing |
| deepDR | Random walk Autoencoder | 10 types of heterogeneous networks | D-D interactions | High accuracy | Low speed |
| GCN-MF | GCN matrix factorization | D-G association Gene features Disease features | D-G interactions | High accuracy | Low speed |
| OWL | Semantic Web technology | PharmGKB FDA approved BCDs | D-D pairs | Text input only | No gold standard testing |
| SLAP | Semantic Link Association | Semantic linked data | D-D interactions | High accuracy | classification only |

Note: CNN: Convolutional Neural Networks; D-D association: drug-disease association; D-D interactions: drug-disease interactions; D-G association: drug-genes association; D-G interactions: drug-genes interactions; D-D pairs: drug-disease pairs.

## 2.1 Drug-disease association prediction based on neural network algorithms

### 2.1.1. Algorithm overview

In a predictive model that uses neural networks to deduce correlations between drugs and illness, the problem is commonly structured as a classification task. This task consists of two primary stages: Feature engineering and categorization. During the feature processing stage, drug and disease features are extracted separately, and then combined into drug-illness feature pairs. During the categorization procedure, a classifier is used to predict and classify the extracted drug-illness features, ultimately producing a classification result [5].

The prediction process for a predictive model for deducing correlations between drugs and disease conditions is illustrated in Figure 1. First, the model computes various characteristics from the drug and disease database and known drug-disease associations, such as chemical structure similarity of drugs, Gaussian interaction contour kernel similarity of drugs and diseases, semantic similarity of diseases, sigmoid kernel similarity of drugs and diseases, and others. Alternatively, the prediction approach may extract latent features of drugs and diseases through autoencoders. Second, the prediction approach fuses Numerous resemblances between drugs and diseases conditions to obtain a comprehensive representation of their characteristics. Next, a designated neural network algorithm is employed by the model to extract extensive features from the drugs and diseases and combine them into drug-disease feature pairs. Ultimately, these interacting pairs are then fed into a classifier for categorization, which yields the likelihood of an association between the drug-disease pair. This can then be used to guide drug relocation efforts.

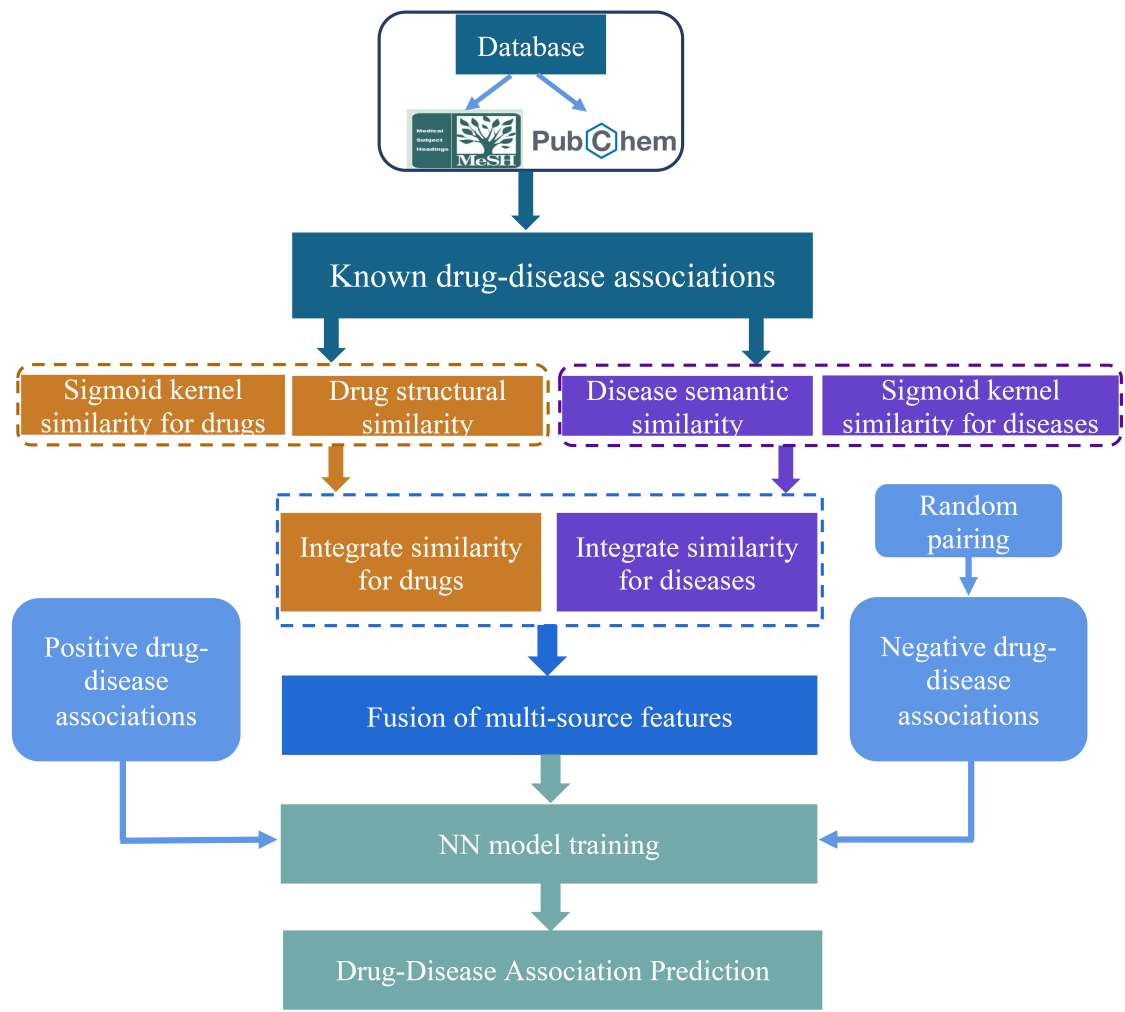

Figure 1 The general procedure of predicting drug-disease associations based on neural network algorithms

Recently, there have been significant progress and breakthroughs in employing neural network algorithms to forecast drug-disease associations. As an example, Jiang *et al.* [16] suggested a model named GIPAE in 2019 which used two novel techniques of a Gaussian interaction contour kernel and autoencoder to predict drug-disease correlations. Autoencoder is utilized to acquire the structural feature depiction of drug fingerprints, and Gaussian interaction contour kernel is employed to compute the resemblance between drugs and diseases. The model integrates diverse kinds of similarity measures, including disease Gaussian interaction profile kernel similarity, drug Gaussian interaction contour kernel similarity, medicinal chemical structure similarity, and disease semantic similarity, to generate comprehensive numeric illustrations of significant disease and drug features. In addition, the fused features are input into the FC neural network to further extract the features of diseases and drugs, and the random forest is selected utilized for classification. The model was validated using various methods, indicating its reliability for predicting drug-disease associations. GIPAE utilizes deep learning technology to extract features and employs an integrated model for prediction. This approach allows for high prediction accuracy to be achieved. In the same year, Jiang *et al.* [17] put forward a sigmoid kernel-based CNN model called SKCNN to predict drug-disease associations. SKCNN employ sigmoid kernel function to construct disease sigmoid kernel similarity and drug sigmoid kernel similarity, and incorporates disease semantic similarity, drug structure similarity. This model combines sigmoid kernels and convolutional neural network technologies to effectively learn representations of drug-disease associations through its hidden layer. The final classification labels are predicted using a random forest classifier. The experimental outcomes indicate that the approach has enhanced the predictive performance. In the case study, it was found that the majority of drugs predicted by SKCNN and the CTD database verified the correlation between obesity and asthma. Jiang *et al.* [18] introduced a model for forecasting drug-disease correlations on a large scale in 2020, which integrates a rotating forest and a sparse autoencoder deep neural network. This approach extracts various features from drugs and diseases, such as Gaussian interaction contours, drug structure similarity, kernel similarity, and disease semantic similarity, to generate a holistic representation of disease and drug characteristics. A sparse autoencoder-based spin forest classifier is then used to anticipate the correlation between diseases and drugs. In comparison to prior methods, the model has significantly improved performance. In 2021, Xuan *et al.* [19] introduced a model called GFPred for predicting drug-disease correlations using fully connected autoencoder (FCA) and attention-based graph convolutional autoencoder (GCA). Utilizing the pre-existing GCA and FCA modules, this approach acquires topological representations of diverse heterogeneous networks and numeric depictions of characteristics drug and disease nodes, separately. At the level of attribute, an attention mechanism is devised to discern the impact of various Characteristics of drug nodes and allocate different weights adjustably. The model deeply integrates the topological representation, attribute representation, and raw attributes of each pair of drug and disease nodes to achieve accurate estimation of their association likelihood. The Graph Convolutional Autoencoder (GCAE) is highly efficient in extracting features between drugs, which allows GFPred to effectively identify similarities between different drugs. The comparison with other approaches proves that this method performs superior to several advanced prediction methods.

In 2020, Jiang *et al.* [18] proposed the SAeRof model, which combines sparse autoencoders and rotating forests to predict drug-disease associations by extracting various similarities, such as Gaussian interaction contour kernel, disease semantics, and drug structure, in order to discover unknown drug-disease interactions. First, the model calculates the Gaussian interaction profile kernel, medicinal chemical structure, and disease semantics between drugs and diseases. The chemical development kit (CDK) is used to calculate the drug similarity, relying on the chemical structures of all drug compounds in SMILES, and is adjusted by a logistic function to obtain drug structural similarity. Then, a drug-weighted network is constructed Founded on established drug-disease associations, where each vertex in the network represents a group of drugs, and an edge is formed between sets of drugs that share a common disease, with the shared disease of a drug pair representing its weight. The ClusterONE algorithm is used to cluster drugs on a disease-sharing network, enhancing the similarity between drugs in the same cluster and obtaining comprehensive drug similarity for similar diseases. Informed by the clustering results, the comprehensive drug similarities $DE$ are calculated. Finally, a sparse autoencoder-based spin forest classifier is suggested to forecast drug-disease associations.

MimMiner was applied to compute disease semantic similarity and develop disease-associated networks based on established drug-disease associations. Within the obtained network, nodes denote diseases, while the weights indicate the frequency of shared drugs among disease. ClusterONE was then employed to cluster diseases in the network, in order to enhance the similarity among diseases within the same cluster. Comprehensive disease similarities $DS$ were subsequently obtained based on the clustering results. Here a drug-disease adjacency matrix $A$ is constructed, which stores verified and unverified drug-disease interactions between drugs $g(j)$ and diseases $d(i)$. The drugs are Depicted by the columns of the matrix, while the diseases are represented by the rows. The i-th column vector of the adjacency matrix $A$ is denoted by a binary vector $V(g(i))$, and the Gaussian interaction profile kernel for drug $g(j)$ and drug $d(i)$ is derived as follows:

$$GE(g(i), g(j)) = \exp(-\theta_g \|V(g(i)) - V(g(j))\|^2) \tag{1}$$

$$\theta_g = \theta_g' / [\frac{1}{nd} \sum_{u=1}^{nd} \|V(g(u))\|^2] \tag{2}$$

Among them, the parameter $\theta_g$ is the adjustable kernel variance, and the original $\theta_g$ parameter is normalized.

The formula for calculating disease Gaussian interaction contour kernel similarity is akin to that applied for computing drug similarity, and is expressed as follows:

$$GD(d(i), d(j)) = \exp(-\theta_d \|V(d(i)) - V(d(j))\|^2) \tag{3}$$

$$\partial_d = \partial_d' / [\frac{1}{md} \sum_{u=1}^{md} \|V(d(u))\|^2] \tag{4}$$

where the binary vector $V(d(i))$ or $V(d(j))$ indicates the affiliation profile by detecting whether $d(i)$ ( or $d(j)$) is linked to each drug and is analogous to the vector in a row of the adjacency matrix A *i*-th (*j*-th). The parameters $\partial_d$ implement the adjustment of the kernel width and the normalization of the original parameters $\partial'_d$. For simplicity, set the $\theta'_g$ and $\partial'_d$ values to 0.5.

Second, the characteristics of drugs and diseases are fused separately. Fill the drug semantic similarity $DE$ in the drug Gaussian interaction contour kernel similarity $GE$ to form the drug similarity matrix $SIM_{drug}$. The drug similarities $SIM_{drug}(g(i), g(j))$ of drugs $g(i)$ and drugs $g(j)$, the drug formulas are as follows:

$$SIM_{drug}(g(i), g(j)) = \begin{cases} GE(g(i), g(j)) & \text{if } g(i) \text{ and } g(j) \text{ has Gaussian} \\ & \text{interaction profile kernel similarity} \\ DE & \text{otherwise} \end{cases} \quad (5)$$

To compute the similarity between diseases, the disease semantic similarity is incorporated into the disease Gaussian interaction distribution kernel similarity formula, which can be written as:

$$SIM_{disease} = \begin{cases} GD(d(i), d(j)) & \text{if } d(i) \text{ and } d(j) \text{ has Gaussian} \\ & \text{interaction profile kernel similarity} \\ DS & \text{otherwise} \end{cases} \quad (6)$$

Then, a sparse autoencoder is utilized to derive the characteristics of drugs and diseases, with a regularization term that induces sparsity introduced to facilitate learning of corresponding sparse features. The cost function is:

$$C_{sparse}(W, b) = C(W, b) + \gamma \sum_{t=1}^{S_2} KL(\rho \| \rho) \quad (7)$$

where $C(W, b)$ is the cost function, $\gamma$ is the weight. After the extraction of features by sparse auto-encoding, principal component analysis (PCA) is utilized to perform dimensional reduction. This process eliminates data redundancy and noise, simplifies the data, and improves the speed of data processing while shortening processing time and reducing processing cost. PCA works by projecting high-dimensional feature vectors onto a lower-dimensional feature space. These newly created orthogonal features, also referred to as principal components, are the eigenvectors of the original high-dimensional feature vectors. Finally, the dimensionality-reduced drug-disease feature vector is inputted into a rotating forest classifier for Categorization and prediction of drug-disease associations.

### 2.1.2. Advantages and disadvantages

One advantage of the algorithm based on neural networks for forecasting drug-disease associations is its ability to extract deeper features of drugs and diseases utilizing different feature extraction methods. This enables the fusion of rich disease and disease drug characteristics in the prediction model, improving its performance. Additionally, the integration of distinct disease and drug similarity information in the neural network-based algorithm can enhance the predictive accuracy of the algorithm. However, this method has certain limitations. The drug and disease data employed may have incomplete characteristic information or unknown relationships, resulting in inaccurate predictions. Moreover, most current neural network-based algorithms only integrate two similarities of drugs and diseases, and it is necessary to further improve the model to integrate more similarities. Furthermore, there is often a significant amount of noise in the drug and disease databases, which can influence the predictive capability of the model. Thus, further improvements to the model are necessary to address this issue.

## 2.2 Matrix-based prediction of drug-disease associations

### 2.2.1 Algorithm overview

The drug-disease association prediction model relying on the matrix algorithm can be broadly classified into two types: matrix completion and matrix decomposition. In matrix completion, the low-rank matrix approximations are used to identify the unrecorded components in the drug-disease Correlation matrix, which can be utilized to detect possible unconfirmed drug-disease relations [20]. The overall prediction procedure is depicted in Figure 2, where the eigenspace of the matrix is constructed to cover the blank items in the correlation matrix for all linear eigenvalues. On the other hand, the matrix decomposition approach takes a single initial matrix and endeavors to derive two additional matrices, which are then multiplied to approximate the input matrix. The prediction process is shown in Figure 3. This technique is similar to finding unobserved associations in the input matrix and is effective for solving prediction problems. Instances of this category of matrix factorization techniques comprise Kernel Bayesian Matrix Factorization (KBMF2K) and Collaborative Matrix Factorization (CMF) [21].

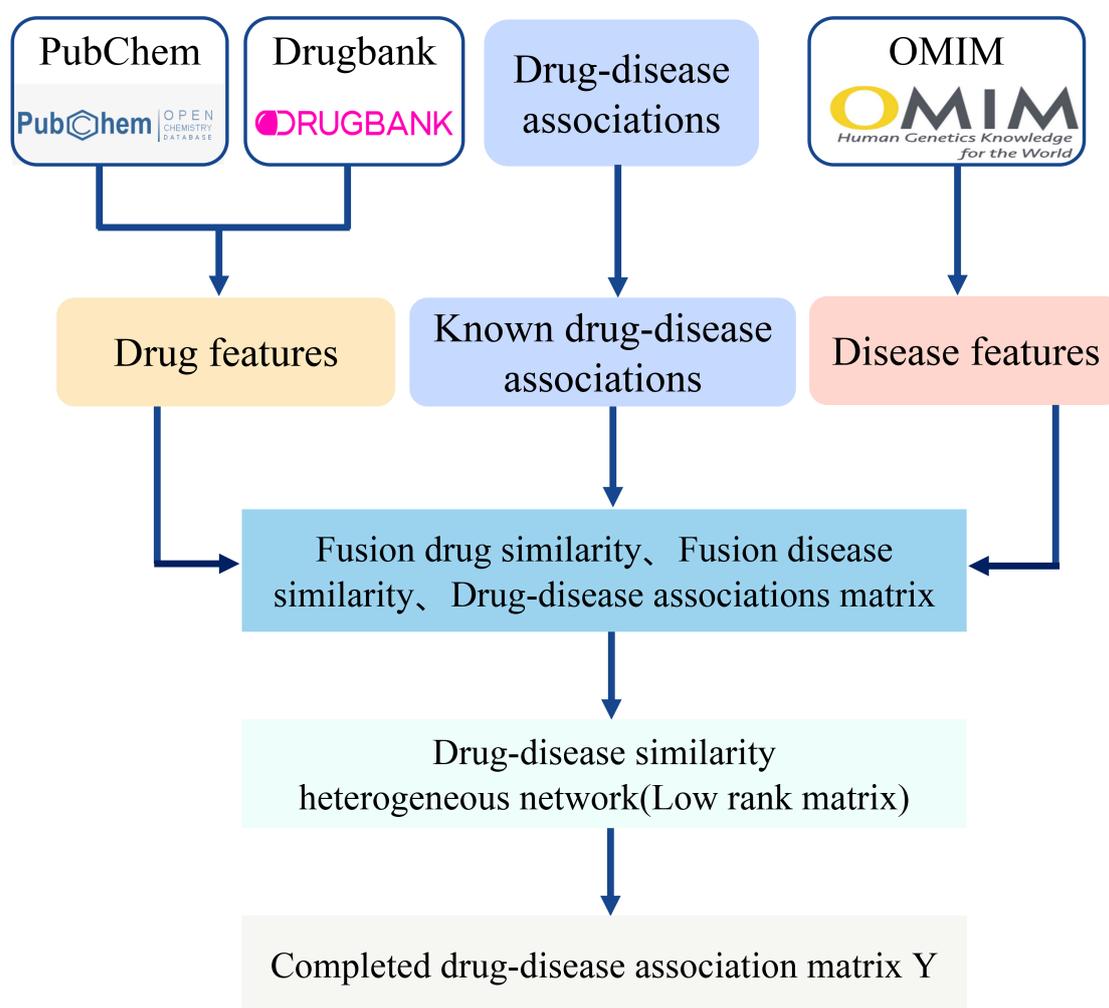

Figure 2 The general process of drug-disease association prediction based on matrix reconstruction algorithm

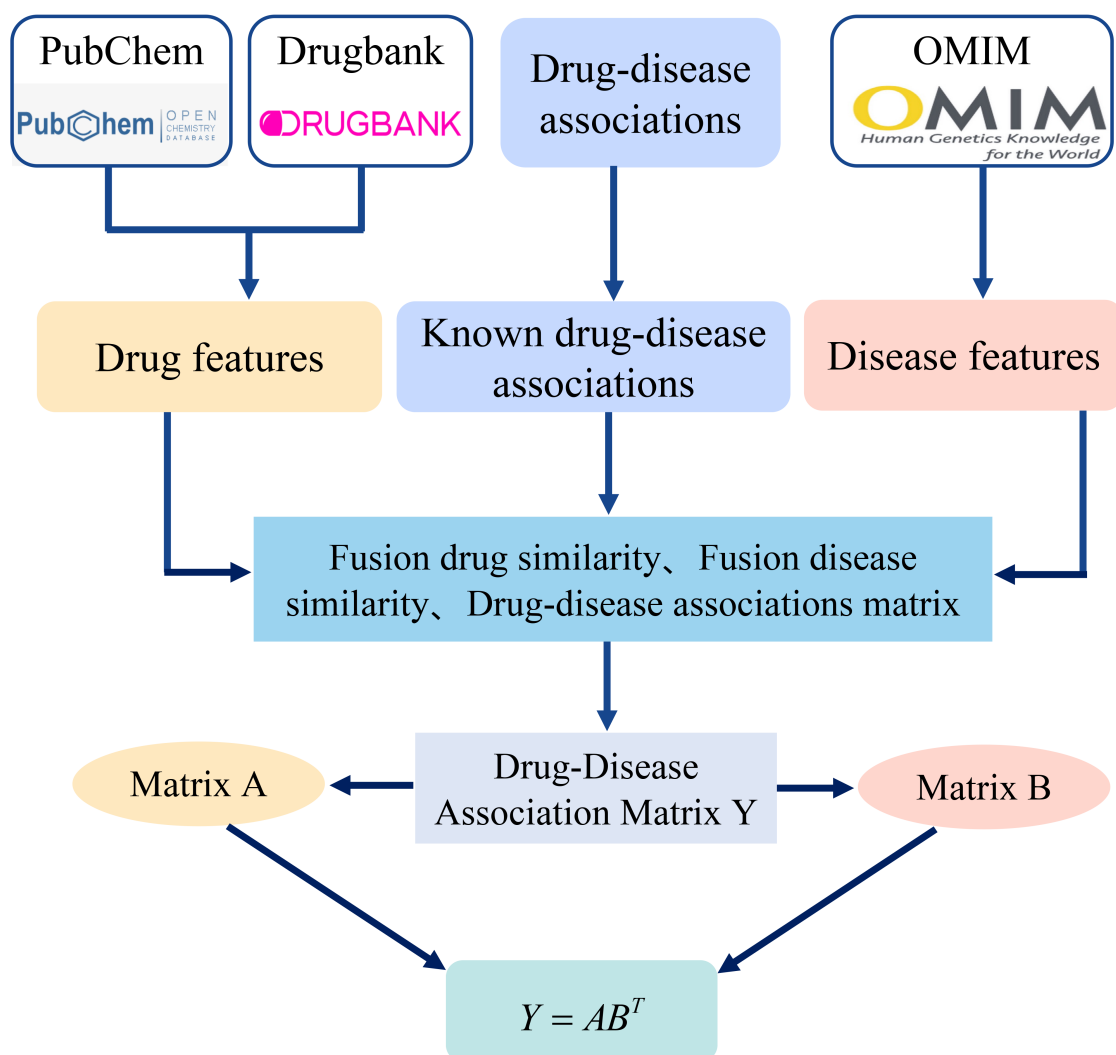

Figure 3 The general process of drug-disease association prediction based on matrix factorization algorithm

In the past few years, numerous scholars have investigated prediction models for drug-disease associations that rely on matrix algorithms. Specifically: Luo *et al.* [20] (2018) recommend a Drug Retargeting Recommendation System (DRRS) that integrates disease-disease, drug-drug, and drug-disease networks to establish a heterogeneous network of drug-disease interactions. A fast and advantageous technology, Singular Value Thresholding (SVT), was employed to predict the scores of drug-disease adjacency matrices for unobserved drug-disease pairs. While each approach has its own advantages for anticipating drug-disease correlations, the current best method is DRRS, as it achieved the greatest AUC score and the most accurate prediction. DRRS can also be utilized to predict drugs with no previously identified disease association. After erasing the known related diseases of a specific drug in the correlation matrix, DRRS can still achieve a better AUC value in predicting the drug-related information. Another method proposed by Cui *et al.* [21] (2019) aims to predict potential drug-disease relationships by combining various similarity matrices. This approach uses the calculated drug network similarity matrix and disease network similarity matrix, in combination with the medicinal chemical similarity matrix and disease semantic similarity matrix, to obtain the drug core and disease core, respectively. To address the issue of unobserved drug-disease associations in the original drug-disease interaction matrix that are missing, the method uses weighted K-nearest known neighbors (WKNKN). Additionally, by introducing the $L_{2,1}$ norm into the CMF technique, the drug-related data and disease-related data are integrated, and the original drug-disease interaction matrix is factorized into two initialized sub-matrices that are subsequently optimized. The ultimate prediction matrix is attained by multiplying the two optimized sub-matrices. The AUC value of this method on some specific datasets is slightly higher than that of DRRS. Huang e*t al.* [22] suggested a multi-task learning (MTL) model based on collaborative matrix factorization, referred to as CMFMTL that not only predicts drug-disease associations but also its corresponding types of associations. In this model, the association matrix for each link type is separately approximated through matrix factorization. The low-dimensional latent representation of drug-disease is shared in the two correlated tasks to achieve the objective of collaborative learning. The CMFMTL model can capture the correlation between the two tasks and effectively leverage all the relevant information to achieve superior accuracy and robustness in performance.

Next, we present a comprehensive account of the DRRS [20] algorithm. The authors believe that the issue for drug repurposing to be conceptualized as a system that identifies novel drugs through the interaction of existing drugs and diseases. The algorithm employs the principle of matrix completion. Under the assumption that the unknown factors associated with drugs and diseases are closely related, the related matrix has a low rank. Subsequently, matrix completion algorithms can fill in the missing values in the disease and drug matrices by building low-dimensional matrix approximations, which can be utilized to identify potential unconfirmed drug and disease interactions.

Initially, a heterogeneous network comprising of drugs and diseases is established. For drugs and drug networks, let $R = \{dr_1, dr_2, ..., dr_m\}$ represent $m$ drugs and the weight of each edge that links two drugs is defined by the pairwise similarity values of their chemical structures. Likewise, for diseases and disease networks, make $D = \{ds_1, ds_2, ..., ds_n\}$ represents a set of n diseases, and each edge that connects two diseases types is assigned a weight based on pairwise phenotypic structural similarity values. A bipartite graph $G(R, D, E)$ is employed to represent the network of drugs and diseases; which $E(G) \subseteq R \times D, E(G) = \{e_{ij}\}$ includes the edge between drug $dr_i$ and disease $ds_j$. If it is assumed that there exists a certain correlation between the drug $dr_i$ and the $ds_j$ disease, the weights of the edges $e_{ij}$ of are initialized to 1, and vice versa 0. By employing the drug and disease correlation network to connect the drug and drug, disease, and disease network to establish a heterogeneous network. The adjacency matrix expression for the heterogeneous network is given as:

$$A = \begin{bmatrix} A_{RR} & A_{RD} \\ A_{RD}^T & A_{DD} \end{bmatrix} \quad (8)$$

In the matrix $A$, the diagonal subarrays $A_{RR}$ and $A_{DD}$ correspond to the affiliation matrices of the drug and disease networks, respectively. Both are dense. The submatrix off the diagonal entries $A_{RD}$ denotes the correlation matrix of the drug and disease network $A_{DR} = A_{RD}^T$, and $A_{RD}^T$ is the transpose of $A_{RD}$. The link matrix of hybrid networks is symmetric and positive semi-definite owing to the bidirectional nature and non-negative weights of the connections in each biological network. Therefore, the eigenvalues of the affiliation matrix A are positive real numbers. The off-diagonal submatrices $A_{RD}$ and $A_{DR}$ exclusively Include the unknown entries, which denote the unobserved associations requiring prediction. Ultimately, the objective of the drug and disease association prediction problem is to complete the missing entries in the adjacency matrix. Then, matrix completion is performed By reducing the sum of the

singular values $A$, the kernel norm $A$, using the relaxation formula as follows:

$$\min \tau \|A^*\|_* + \frac{1}{2}\|A^*\|_F^2$$
$$s.t. P_\Omega(A^*) = P_\Omega(A)$$
(9)

Beginning at $Y^{(0)} = \lceil \tau/(\delta\|P_\Omega(A)\|)\rceil \delta P_\Omega(A)$, SVT generates a set of matrices $(X^{(i+1)}, Y^{(i+1)})$ to reconstruct Uzawa's algorithm or linearized Bregman iteration by the following formula, the particular equation is given by:

$$\begin{cases} X^{(i+1)} = D_\tau(Y^{(i)}) \\ Y^{(i+1)} = Y^{(i)} + \delta P_\Omega(A - X^{i+1}) \end{cases}$$
(10)

And because the size of the iterative step is determined to $(m+n)/\sqrt{|\Omega|}$, the SVT operator $D_\tau(.)$ represents a soft threshold operator, and the equation can be modified as:

$$D_\tau(Y^{(i)}) = \sum_{j=1}^{\sigma_j^{(i)} \geq \tau} (\sigma_j^{(i)} - \tau) u_j^{(i)} v_j^{(i)T}$$
(11)

where $\sigma_j$ includes singular values greater than $\tau$, $u_j$ and $v_j$ denote the singular vectors on the left and right of $D_\tau(.)$, respectively.

When performing SVT for matrix reconstruction, the singular values $D_\tau(.)$ of $Y^{(i)}$ that exceed the computed threshold $\tau$ need to be estimated at each iteration step. This can be acquired directly by calculating the SVT of $Y^{(i)}$, and then reducing it by choosing singular values $\tau$ larger than and its associated singular vector. Nevertheless, performing the complete numerical computation of the singular value decomposition for adjacency matrices of large heterogeneous networks is often computationally and memory intensive. In reality, during the iteration process of SVT, $D_\tau(.)$ only the singular values $Y^{(i)}$ in are larger than $\tau$ are involved. This facilitates the use of a rapid singular value $b$ decomposition algorithm to estimate the significant singular values of concern, which enhances the computational speed of the matrix completion algorithm. A rank revealing random SVD algorithm (R3SVD) is proposed $Y^{(i)}$ by projecting to a small Gaussian matrix and using iterative power algorithm. R3SVD Constructs a low-rank QB factorization by utilizing incremental orthogonal Gaussian projections, which is then used to obtain a low-rank SVD. The stochastic SVD (R4SVD) algorithm, which extends R3SVD to cyclic rank, enhances the computational capability of the SVT algorithm by utilizing regular vectors derived from prior iterations. Here, the R4SVD algorithm is integrated into DRRS for fast computation $D_\tau(.)$. A speedy execution of the SVT algorithm utilizing R4SVD, called SVT-R4SVD, is used to perform matrix completion in the DRRS method. Finally, drug repurposing is achieved by verifying the new association relationships in the completed matrix.

### 2.2.2 Advantages and disadvantages

Matrix-based drug-disease association prediction usually uses traditional computing methods, and association prediction is often transformed into an optimization problem. Our goal is how to solve it efficiently. Compared with the neural network algorithm, the matrix-based method has faster model training efficiency and better predictive performance. Its shortcoming is that the ability to further extract the characteristics of drugs and diseases is not as good as that of neural networks.

The main advantage of the drug-disease association prediction technique built upon the matrix completion algorithm is that it can take into account all the dominant eigenvalues of the adjacency matrix and their related eigenvectors. However, the disadvantage is that the measurement of data sparsity and similarity in the dataset it uses may affect the prediction effectiveness. To address this issue, it may be indispensable to collect and integrate more pertinent linked information from multiple databases or literature.

The advantage of the drug-disease interaction prediction method founded on matrix factorization algorithm is that it can integrate network information regarding drugs and diseases, consider various similarity information in the prediction algorithm, and achieve better prediction performance. However, its limitation is that model training is time-consuming and takes longer to train than other methods.

## 2.3 Prediction of drug-disease association based on recommendation algorithm

### 2.3.1 Algorithm overview

In the research of drug relocation based on recommendation algorithm, the recommendation method of collaborative filtering (CF) in the recommendation model is the most widely used recommendation algorithm. Recommendation algorithms in CF can be broadly categorized into three groups: item-based, user-based and model-based. User-based recommendation approach, that is, finding similar neighbor users through common tastes and preferences, K-neighbor algorithm [23, 24]. For example, if your friend likes a certain movie, you may also like it. Item-based recommendation algorithms find similarities between items and recommend similar items. For example, if your preference is for item A and there exist certain similarities between item A and item C, it is reasonable to assume that you may also have a liking for item C. Model-based recommendation algorithms develop a recommendation system founded on the user interest data of the sample and then generate proposals according to current user liking information. CF algorithms utilize similar correlations between users or items to make recommendations based on this information.

In the field of drug retargeting, collaborative filtering algorithms assume that similar drugs or diseases may have a common indication or drug candidate. This is predicted by aggregating previously known disease scores for similar drugs to target the drug score for a specific disease, or by searching to help with the anticipation of related diseases of drug candidates against the target illness.

CF is a successful recommendation algorithm in recommender systems. Essentially, it utilizes a user's past purchasing, rating, browsing, and other recorded information to recommend information to the user or predict their interests and preferences, thereby achieving personalized recommendation results [25, 26]. In the user-based collaborative filtering model for drug relocation, drugs are typically treated as users and diseases as items. The main idea is to capture the correlation between existing drugs, since similar drugs often share similar indications [1, 27, 28]. Therefore, in drug relocation based on collaborative filtering, a newly defined similarity measure method is first used to calculate the degree of convergence between drugs and establish a similarity matrix of responses. Then, a new collaborative filtering model is constructed to estimate the association between drug combinations and illnesses. Finally, the estimated probability of the drug's efficacy on the disease is calculated [24]. The general prediction process is shown in Figure 4.

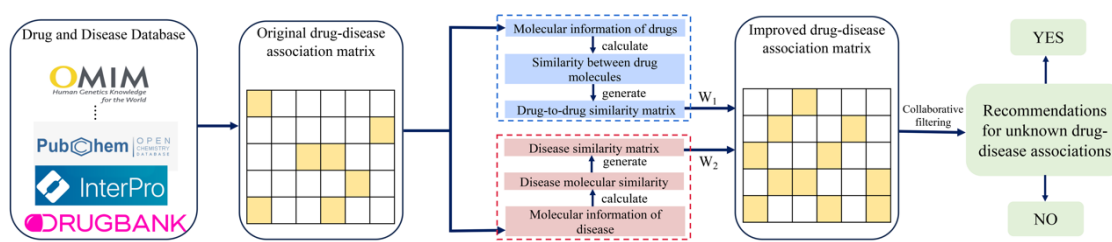

Figure 4 General process of anticipating the link between drugs and illnesses using recommendation algorithm

In the current years, researchers have made significant advancements in exploring drug-disease linkage prediction models using recommendation algorithms. To demonstrate, Lin *et al.* [24] proposed a drug relocation algorithm using CF in 2015. This method involves constructing a drug-disease correlation matrix using gathering descriptive details on drugs and diseases, including indications for diseases and related side effects. A measure is then used to describe the degree of convergence between drugs, which helps to measure the similarity of different drugs in terms of indications and side effects. Based on this measure, a similarity matrix of responses is created. Using the constructed collaborative filtering model, the link between the medication and the illness is predicted, and the prediction score of the drug on the disease is calculated. The experiments demonstrate that this method can not only boost the predictive capability of the system, but also effectively identify therapeutic drug-disease combinations. Zhang *et al.* [28] presented a computational drug relocation model built on multi-source fusion collaborative filtering in 2017. In this method, multiple data sources, such as drug-disease associations, medicinal target proteins and medicinal chemical structures, are integrated to derive similarity matrices for drugs and diseases. CF is then used to obtain multiple estimated probabilities based on different similarity matrices, and the weight learning method is applied to integrate these scores into the optimization results, completing the task of drug repositioning. The authors compared multiple data sources, including DDAS (drug-disease association), TAPR (target protein), and CHST (chemical structure), and discovered that DDAS was the most critical data source based on their ablation experiments. The trial results demonstrate that this framework is not only outperforms in diverse evaluation metrics, but also effective in identifying the potential of drug treatments. Among the ten medications for stroke treatment forecasted by the model, three drugs have already undergone clinical trials.

### 2.3.2  Advantages and disadvantages

The advantages of the collaborative filtering and recommendation algorithm are evident. This model is very adaptable and does not demand extensive expertise in the relevant data domain. The project is simple to implement, and the effect is also quite good. However, it also has certain limitations, such as the "cold start" problem. When there is no drug-related information available, it cannot recommend treatments for new drugs. Additionally, it does not consider differences in various scenarios, such as symptoms of unforeseen diseases.

## 2.4 Methods based on text mining and semantic reasoning

### 2.4.1  Algorithm overview

Text mining methods are commonly employed in drug retargeting studies to search for data pertaining to a specific disease, gene, or drug. The retrieved data is then analyzed to identify related entities or knowledge using co-occurrence information or natural language processing-based classification. In cases where gene G is associated with disease D, and drug R is linked to gene G, there is a possibility that drug R may also be linked to disease D. Typically, text mining generally involves four key stages: Knowledge Discovery (KD), Information Extraction (IE), Information Retrieval (IR), and Entity Recognition (NER)[29]. These steps enable researchers to efficiently extract and analyze large volumes of data from various sources, facilitating the identification of novel drug-disease associations. It is essential to ensure that the language is polished, and grammatical errors are minimized, while minimizing redundancy in the text.

Text mining is a valuable tool used to explore links between drugs, diseases, and genes, investigate gene-gene associations, and construct diverse networks of diseases, drugs and genes. In a recent study, Li *et al.* [30] introduced an innovative method that incorporates data from mining of text-based literature and networks of protein interactions to construct drug-protein correlation maps for particular diseases. As an illustration, the authors focused on Alzheimer's disease (AD) and demonstrated that their approach surpasses traditional information retrieval systems and drug target databases. The study also identified two currently available medications as potential candidates for AD treatment.

In contrast to conventional literature mining strategies that construct biological networks constructed from the co-occurrence of biological structures, Tari *et al.* [31] proposed a unique method to biological network construction that differs from traditional text mining methods. Their approach takes into interaction type orientations, account interaction types, and drug mechanism representations. Using text mining, the authors gathered information from openly accessible sources, which led to the creation of a group of logical propositions. These facts were then utilized to develop an automated inference model. This model, based on logical rules that represent the mechanistic properties of the drug, is capable of identifying the therapeutic potential of existing drugs and new indications. In a similar vein, Rastegar Mojarad *et al.* [32] utilized text mining data to detect semantic predictions of gene-disease relationships and drug-gene, which were subsequently employed to formulate a series of possible drug-illness pairs. Based on the experimental findings, a significant proportion of the drug-illness pairs with high projected scores are present in the Comparative Toxicogenomics Database (CTD) when compared with the predicted samples. The authors determined that by prioritizing these pairs using the predicates that link drug-gene and gene-disease pairs, a synthesis of drug-gene and gene-disease predicates could identify illness among the drug-disease pairs with the highest predicted scores as potential candidates for drug repurposing. Brown *et al.* [33] developed a drug repurposing text data analysis system accessible through the web, which clusters drugs based on their shared indications to identify both known and new drug indications. The authors presented an end-to-end case study for metformin to demonstrate the effectiveness of their model. Meanwhile, Papanikolaou *et al.* [34] utilized text mining to recognize drug co-occurrences in the DrugBank database to detect biological components (like proteins, diseases and genes) in the drug library indication, description, drug action, and mode of operation text fields, the authors employed named entity recognition (NER) techniques. After eliminating unimportant terms, they created binary vectors to represent each drug library record, and clustered drug library records utilizing various clustering techniques and similarity metrics. By utilizing this method, it is possible to identify novel drug-drug associations, which could be beneficial in drug repurposing scenarios.

Recently, Zeng *et al.* [35] recently proposed a neural network-driven method for identifying prospective drug-disease interrelationships by building ten heterogeneous networks using data retrieved from various public sources. Their method outperformed traditional methods in identifying new connections between drugs and diseases and suggested prospective drug repurposing options for Alzheimer's and Parkinson's disease. The deepDR model was compared with the selected baseline method on cross-validation and external validation groups, and the findings indicated that deepDR surpassed the e baseline with higher AUROC values. Meanwhile, Han *et al.* [36] applied mining of OMIM phenotypes using text analysis to develop phenotype networks and utilized graph convolutional neural networks (GCNNs) to detect associations between diseases and genes through emphasizing nonlinear correlations between diseases and genes. The authors gathered data from the Online Mendelian Inheritance in Man (OMIM) database, and their proposed method achieved optimal values for almost all metrics in each fold of the three-fold cross-validation.

Furthermore, semantic technologies facilitate the amalgamation of diverse data repositories and the identification of novel drug indications. For example, Zhu *et al.* [37] designed an ontology that represents SNPs, genes, drugs, pathways and diseases associated with FDA-approved breast cancer drugs. They employed an ontology-based knowledge base to deduce novel drug-illness pairs. Empirical findings demonstrate that Semantic Web technology can bring better performance for the prediction of new indications of breast cancer drugs. Similarly, Chen *et al.* [38] constructed a mathematical model that evaluates drug-target connections using a semantically interlinked network comprising protein, diseases, drugs targets, compounds, pathways, side effects, and their relationships. The model assesses the structure and meaning of subgraphs that link drugs and targets and recognizes comparable drug-drug pairs from distinct disease regions, which could suggest potential drug repositioning prospects. Indirect drug-target pairs can also be identified, such as drugs that can modify gene expression levels, although they may not be as potent as target pairs known to interact directly.

### 2.4.2 Advantages and disadvantages

The advantage of utilizing text mining and semantic reasoning methods is that a large repository of information regarding diseases, drugs, and genes can be accessed, coupled with the fast-paced development of research literature in the fields of biology, biomedicine, and medicine. Data mining techniques can be applied to uncover a wealth of information that is otherwise hidden in the literature. Furthermore, information from various data sources can be easily integrated, accelerating the prediction of the therapeutic potential of existing drugs and new treatable diseases. This offers a new approach to drug-disease association prediction. Nonetheless, there exist certain limitations to these approaches. The prediction of drug-disease associations cannot be completely achieved only through text mining and semantic reasoning methods, and it needs to be combined with other computing methods to obtain more accurate prediction results.

## 3. Prediction performance comparison and discussion

To effectively illustrate the predictive performance of models for predicting drug-disease associations utilizing different algorithms, this review selects one classic prediction model from each category of algorithms and tests its predictive performance using the same dataset.

The dataset utilized in this study is compiled by Luo *et al.* [39, 40] and contains drug and disease association information. As shown in Table 2, the dataset comprises drugs (663), diseases (409), and validated drug-disease associations (2532), which serve as the gold standard dataset (hereafter referred to as "Cdatasets"). The DrugBank database provided the drug-related data, which contains extensive drug-related information [41]. The disease information was sourced from the OMIM (Online Mendelian Inheritance in Man) database, renowned for its focus on genetic diseases and comprehensive coverage of textual information, relevant reference information, and sequence records [42]. The chemical structure of the drug, also known as the drug fingerprint, was obtained from the PubChem database [43]. In this study, negative samples were randomly generated from unlabeled drug-disease pairs to match the positive samples [16].

Table 2 Cdatasets dataset

| Dataset  | drug | disease | associate |
|----------|------|---------|-----------|
| Cdataset | 663  | 409     | 2532      |

### 3.1 Comparison of AUC and AUPR values

The representative algorithms selected for predicting drug-disease associations were compared using ten-fold cross-validation and two evaluation metrics: AUC and

AUPR. The comparison included neural network-based (2020SAeRof, 2019SKCNN)[17, 18], matrix-based (2019L2, 1-CMF, 2020CMFMTL)[21, 22], and recommendation algorithm (2019CFNBC, 2020HCFMDA) [44, 45]. The Cdatasets, introduced earlier, were used for this purpose. Figure 5 displays the results of different algorithms' predictions on this dataset. Specifically, Figure 5A compares the algorithms' AUC values, while Figure 5B compares their AUPR values.

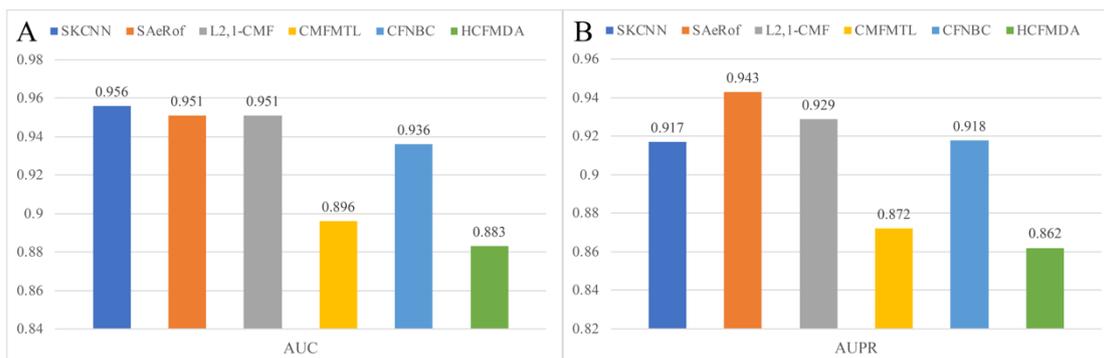

Figure 5 Comparison of different algorithms based on AUC and AUPR values

Referring to the prediction results presented in Figure 5, it is evident that the drug-disease association prediction models utilizing neural networks exhibit better performance compared to the matrix and recommendation-based algorithms to a certain extent, as evident from the AUC scores. However, it is crucial to acknowledge that the prediction ability of a certain model cannot be solely judged based on this aspect and needs to be further assessed in conjunction with other performance evaluation indicators, such as the model's availability and reliability.

## 3.2 Dataset imbalance

In drug-disease association prediction research, a common challenge is the imbalance between positive and negative samples. Due to the rarity or insufficient study of certain drug-disease associations, positive samples (known drug-disease associations) are significantly outnumbered by negative samples (unknown drug-disease associations). Data imbalances can impact the training and performance evaluation of prediction models for drug-disease associations. In these cases, models might tend to predict negative samples, leading to low sensitivity and a high false-negative rate.

Therefore, in drug-disease association prediction research, appropriate data processing and algorithm adjustments are needed to address the data imbalance issue and ensure the predictive performance of the models. Several common approaches are typically employed when facing imbalanced positive and negative samples in drug-disease association prediction research:

1. Resampling: This method balances the dataset by either oversampling the minority-class samples or undersampling the majority-class samples. Common resampling techniques include oversampling and under sampling. Oversampling methods involve augmenting the quantity of minority-class samples through replication or generation, while under sampling methods decrease the number of majority-class samples by removal.
2. Class weighting: By assigning different weights to different classes, minority class samples are given higher importance. Using class weights during model training balances the impact of different classes, ensuring that the model focuses more on minority class samples.
3. Synthetic sample generation: This method utilizes certain generation models (such as synthetic minority oversampling technique (SMOTE), ADASYN, etc.) to create synthetic samples based on existing minority-class samples. These synthetic samples increase the diversity of the minority class in the dataset, thereby improving model performance.
4. Threshold adjustment: In model prediction results, adjusting the classification threshold can balance the model's performance on positive and negative samples. By adjusting the threshold, more emphasis can be placed on the prediction of minority-class samples, thereby improving the model's recall or specificity.
5. Ensemble methods: Ensemble methods combine the predictions of multiple classifiers to obtain the final prediction result through voting or weighted averaging. This approach can alleviate the impact of positive and negative sample imbalances and improve model performance.

## 3.3 Feature fusion

Regardless of the approach used to predict drug-disease association, it is necessary to extract and fuse the pertinent feature information of drugs and diseases in the model. For feature extraction in the neural network-based prediction model, autoencoders such as variational autoencoders, stacked autoencoders, and sparse autoencoders are commonly used for feature extraction. This approach not only obtains rich feature information of drugs and diseases, but also mitigates the impact of data noise on the prediction model to some extent. Additionally, convolutional neural networks, graph convolutional neural networks, and other methods can also be used to fuse different feature information of drugs and diseases, which can improve the prediction performance of the model. In this prediction model, a classifier is usually combined for the final classification prediction. Different classifiers can bring about significant differences in prediction performance, so there is also great room for optimizing the model's prediction performance, and the model can be continuously optimized. Matrix decomposition and matrix completion-based prediction models incorporate the matrix decomposition method, which combines latent semantics and machine learning features to explore deeper drugs-diseases association. Therefore, the prediction accuracy is notably high, surpassing that of neighborhood-based collaborative filtering and content-based recommendation algorithms. When using a content-based recommendation algorithm, it becomes possible to effectively model the physical and chemical properties of drugs. To achieve better recommendation accuracy, researchers can increase the dimension of disease attributes within the algorithm. Nevertheless, its shortcomings are that effectively obtaining more data poses a challenging task when the disease attributes are limited, and the measurement standard for disease similarity only considers its own properties, which has a certain one-sidedness.

### 3.4 Scalability

One aspect that attracts researchers to predictive models based on neural network algorithms is the flexibility of their architecture, which allows for the development of single-task or multi-task models for the identification of potential therapeutic applications and the prediction of drug-disease association. While neural network methods have undoubtedly been instrumental in developing emerging models for drug repositioning, it is essential to acknowledge that they also have certain limitations. Deep neural network models often necessitate substantial time and effort for proper adjustment to the training data used. Moreover, the complexity of selecting the appropriate technique or similarity measure for each dataset within the deep neural network layers is heavily reliant on the distinct characteristics of the dataset itself. Matrix factorization-based methods are highly scalable, and improved matrix factorization methods such as SVD++ and TimeSVD can easily add other elements to drug and disease feature vectors. Additionally, leveraging drug and disease attribute information further boosts the model's predictive performance. Nonetheless, it also has certain limitations, as it requires mapping drugs and diseases to latent factor spaces, making these latent features not easily explainable by real-world concepts, which sometimes leads to poor model interpretability. In drug-disease association prediction models based on recommendation algorithms, collaborative filtering recommendation algorithms are often used, which often have good scalability, that is, the ability to recommend new information and can discover content that is not similar in content, thus recommending potential therapeutic diseases for new drug.

## 4. Current challenges and future prospects

Each method mentioned for predicting drug-disease associations has its own strengths and limitations. To achieve better results, it is sometimes necessary to combine these methods. By doing so, we can leverage each method's strengths and weaknesses and integrate their advantages to improve prediction accuracy. For instance, Wanget *et al.* [46] employed a combination of information including medicinal phenotype data, target protein sequences, and chemical structures, and utilizing machine learning algorithms to predict drug-disease relationships. They also conducted a network analysis of drug-disease relationships. Similarly, Gottliebet *et al.* [47] integrated multi-omics data, including phenotypic data, drug side effects, chemical structures, target protein interactions, and drug target protein sequences, to enhance their predictions of drug-disease associations. They calculated target protein distances through network analysis, identified disease phenotypes through text mining, and applied machine learning algorithms to classify drug-disease associations as true or false based on this comprehensive set of data. The integration of these methods has consistently demonstrated enhanced performance in terms of sensitivity and specificity, surpassing the performance of individual methods. This indicates that the integration of these methods holds significant promise in enhancing drug-disease association predictions.

In the present studies focusing on the prediction of drug-disease associations, it is essential to develop feature extraction methods that can extract more comprehensive drug and disease feature information. Moreover, it is necessary to develop feature fusion methods that can integrate multiple feature information, enabling the inclusion of more drug and disease information in predictive models, ultimately leading to improved prediction accuracy. Most existing drug-disease association prediction methods employ shallow models. However, the relationship between drugs and diseases is nonlinear and complex. Shallow models struggle to capture these intricate relationships, thus hindering their ability to mine advanced levels of data. Therefore, it is imperative to develop models capable of capturing the intricate representations of drug-disease associations to enhance the prediction of drug indications. Furthermore, it is crucial to conduct further research on prediction algorithms and integrate multiple methods to maximize their advantages while minimizing their limitations.

Although various computational methods for predicting drug-disease associations have been developed, there remain significant challenges that must be addressed in this research field. The following are some of the existing challenges:

1) **Data scarcity and imbalance:** Due to the rarity or limited research on certain drug-disease associations, the number of known drug-disease associations (positive examples) is significantly limited when compared to the vast pool of unknown drug-disease associations (negative examples). This leads to data scarcity and imbalance, posing challenges for model training and performance evaluation.

2) **Diversity and complexity:** The diversity and complexity of drugs and diseases make it challenging to predict their associations. Drugs can have multiple mechanisms of action and targets, while diseases can involve multiple biological processes and pathways. This complexity makes it challenging to build accurate prediction models.

3) **Lack of standardized data and shared resources:** Relevant data on drugs and diseases is often scattered across different databases and literature, lacking standardization and unified formats. Additionally, many data sources are still not widely shared, limiting researchers' ability to develop and validate prediction models.

4) **Validation of unknown drug-disease associations:** Due to constraints in time and resources, the validation of a substantial number of unknown drug-disease associations presents significant challenges. Therefore, finding effective ways to validate the accuracy and reliability of prediction models remains a challenge.

5) **Interpretability and explainability:** The interpretability and explainability of drug-disease association prediction models are crucial. In clinical practice, healthcare professionals and researchers need to understand the predictions made by the models and be able to explain the reasons behind them. Therefore, constructing models with high interpretability and explainability is a challenge.

Overcoming these challenges requires interdisciplinary collaboration, including the availability of rich data resources, improved prediction algorithms and methods, better data standardization and sharing mechanisms, as well as a focus on interpretable and explainable model design.

The dug-disease association (DDA) is a crucial area of research because it can reveal the potential efficacy of a drug in treating a specific disease or condition. However, experimentally validated DDAs are still scarce. Previous evidence suggests that the integration of diverse biological data sources can facilitate the discovery of novel DDAs. Nonetheless, integrating such data to determine the optimal drug for treating a specific disease, leveraging the drug-disease coupling mechanism, remains a significant challenge. Despite the numerous models proposed by researchers for predicting drug-disease associations and facilitating drug repositioning, efficiently extracting DDA information remains a persistent challenge. Examining the intricate correlations between drugs and diseases by delving into the microscopic perspective of intracellular biomolecules can offer novel insights into the mechanisms of diseases. Overall, continuing the exploration of innovative approaches to predict DDAs and identify effective treatments for diseases is crucial. Developing efficient techniques for extracting DDA information from various biological data sources is critical for advancing drug discovery research. By understanding the intricate mechanisms underlying DDAs, researchers can potentially develop more effective therapies to improve patient outcomes.

## Acknowledgments

This work was supported by the Natural Science Foundation of China (No. 62072353, No. 62272065, and No. 62202081).

## Conflicts of Interest

All authors declare no conflict of interest.